\documentclass[aps,floatfix,showpacs,twocolumn,superscriptaddress]{revtex4} 
\usepackage{graphicx} 
\begin{document} 
%\preprint{\parbox[t]{45mm}{\small ANL-PHY-10520-TH-2003\\% 
%                                  KSU\\}} 
% 
\title{Analysis of a quenched lattice-QCD dressed-quark propagator}

\author{M.S.\ Bhagwat} 
\affiliation{Center for Nuclear Research, Department of Physics, 
             Kent State University, Kent, Ohio 44242 U.S.A.} 
\author{M.A.\ Pichowsky} 
\affiliation{Center for Nuclear Research, Department of Physics, 
             Kent State University, Kent, Ohio 44242 U.S.A.} 
\author{C.D.\ Roberts} 
\affiliation{Physics Division, Argonne National Laboratory, 
             Argonne, IL 60439-4843 U.S.A.} 
\author{P.C.\ Tandy} 
\affiliation{Center for Nuclear Research, Department of Physics, 
             Kent State University, Kent, Ohio 44242 U.S.A.} 
%\thanks{} 
\date{\today} 
%------------------------------------------------------------------------ 
\begin{abstract} 
Quenched lattice-QCD data on the dressed-quark Schwinger function can be
correlated with dressed-gluon data via a rainbow gap equation so long as that
equation's kernel possesses enhancement at infrared momenta above that
exhibited by the gluon alone.  The required enhancement can be ascribed to a
dressing of the quark-gluon vertex.  The solutions of the rainbow gap
equation exhibit dynamical chiral symmetry breaking and are consistent with
confinement.  The gap equation and related, symmetry-preserving ladder
Bethe-Salpeter equation yield estimates for chiral and physical pion
observables that suggest these quantities are materially underestimated in
the quenched theory: $|\langle \bar q q \rangle|$ by a factor of two and
$f_\pi$ by $30$\%.
\end{abstract} 
%------------------------------------------------------------------------ 
% \keywords{} 
\maketitle 
%------------------------------------------------------------------------ 
\section{Introduction \label{Sec:Intro}} 
It is a longstanding prediction that the Schwinger functions which
characterise the propagation of QCD's elementary excitations: gluons, ghosts
and quarks, are strongly modified at infrared momentum scales, namely,
spacelike momenta $k^2 \lesssim 2\,$GeV$^2$
\cite{cdragw,bastirev,reinhardrev}.  Indeed, this property of asymptotically
free theories was elucidated in Refs.\ \cite{lane,politzer} and could be
anticipated from studies of strong coupling QED \cite{bjw}.  Such
momentum-dependent dressing is a fundamental feature of strong QCD that is
observable in hadronic phenomena \cite{mrrev}.  For example: it is the
mechanism by which the current-quark mass evolves to assume the scale of a
constituent-quark mass at infrared momenta, and thereby that dynamical chiral
symmetry breaking (DCSB) is exhibited; and it may also provide an
understanding of confinement, as we canvass in Sec.\
\protect\ref{sec:confinement}.  These are keystones of hadron physics
\cite{nsac}.

Numerical simulations of lattice-QCD provide direct access to QCD's Schwinger
functions, and recent studies of the quenched theory yield dressed-gluon
\cite{Leinweber:1998uu,latticegluon1,latticegluon2,latticegluonSU2} and
-quark \cite{bowman1,bowman2} two-point functions (``propagators'') that are
in semi-quantitative agreement with Dyson-Schwinger equation (DSE)
calculations
\cite{dsegluon1,dsegluon2,mr97,mt99,petersummary,FischerAlkofer03,zwanziger}.
However, these dressed-gluon and -quark propagators are not obviously
consistent with each other in the following sense: use of the lattice
dressed-gluon two-point function as the sole basis for the kernel of QCD's
gap equation cannot yield the lattice dressed-quark propagator without a
material infrared modification (enhancement) of the dressed-quark-gluon
vertex \cite{fred1,hawes,pc91}.  Fortunately, just such behaviour can be
understood to arise owing to multiplicative renormalisability of the gap
equation \cite{bloch,raya} and is observed in lattice estimates of this
three-point function \cite{latticevertex1,latticevertex2,latticevertex3}.

Herein we elucidate these points using a concrete model for the gap
equation's kernel.  Naturally, since the ultraviolet behaviour of this kernel
is fixed by perturbative QCD and hence model-independent, our study will
focus on and expose aspects of the infrared behaviour of the Schwinger
functions described above.

Furthermore, as indicated at the outset, DCSB is encoded in the chiral-limit
behaviour of the dressed-quark propagator.  However, contemporary lattice-QCD
simulations are restricted to current-quark masses that are too large for
unambiguous statements to be made about the magnitude of this effect.  With a
well-constrained model for the gap equation's kernel it is straightforward to
calculate the dressed-quark propagator in the chiral limit.  Hence our
analysis will also provide an informed estimate of the chiral limit behaviour
of the lattice results.

Finally, we consider two additional questions; namely, how do Schwinger
functions obtained in simulations of quenched lattice-QCD differ from those
in full QCD, and can that difference be used to estimate the effect of
quenching on physical observables?  Our model for the gap equation's kernel
provides a foundation from which we believe these problems can fruitfully be
addressed.

The article is organised as follows.  In Sec.\ \ref{Sec:DSE} we review the
gap equation, the form of its solution and the nature of its kernel.  Section
\ref{Anallattice} describes the construction of a model for the gap
equation's kernel that correlates lattice results for the dressed-gluon and
-quark two point functions, and explains aspects of the dressed-quark
function obtained therewith.  The kernel is exploited further in Sec.\
\ref{Sec:Results}, wherein it provides the basis for calculating informed
estimates of observable quantities in quenched-QCD.  Section
\ref{Sec:Conclusion} is an epilogue.

%------------------------------------------------------------------------ 
\section{Gap Equation\label{Sec:DSE} } 
The renormalised dressed-quark propagator, $S(p)$, is the solution of the
DSE
\begin{equation} 
S^{-1}(p)\!=\!Z_2(\zeta,\Lambda) \, i\, \gamma \cdot p 
    + Z_4(\zeta,\Lambda)\, m(\zeta) 
 + \Sigma'(p,\Lambda), 
    \label{quarkdse} 
\end{equation} 
wherein the dressed-quark self-energy is$^{\,1}$\footnotetext[1]{We use a
Euclidean metric wherewith the scalar product of two four vectors is
\mbox{$a\cdot b=\sum_{i=1}^4 a_i b_i$}, and Hermitian Dirac-$\gamma$ matrices
that obey \mbox{$\{\gamma_\mu,\gamma_\nu\} = 2\delta_{\mu\nu}$}.}
\begin{equation} 
\Sigma'(p,\Lambda) = Z_1(\zeta,\Lambda) \!  \int^\Lambda_q \!\!\!
 g^2D_{\mu\nu}(p-q) \, \frac{\lambda^i}{2}\gamma_\mu \, S(q) \,
 \Gamma^i_\nu(q,p).  \label{quarkSelf}
\end{equation} 
Here $D_{\mu\nu}(k)$ is the renormalised dressed gluon propagator;
$\Gamma^i_\nu(q,p)$ is the renormalised dressed quark-gluon vertex; and
$Z_1(\zeta,\Lambda)$, $Z_2(\zeta,\Lambda)$ and $Z_4(\zeta,\Lambda)$ are,
respectively, Lagrangian renormalisation constants for the quark-gluon
vertex, quark wave function and current-quark mass.
 
In Eq.\ (\ref{quarkSelf}), \mbox{$\int^\Lambda_q := \int^\Lambda d^4
q/(2\pi)^4$} denotes a translationally invariant ultraviolet regularisation
of the momentum space integral, with regularisation mass-scale $\Lambda$.  In
practice a Pauli-Villars scheme is used, in which an additional factor of
$\Lambda^2/(\Lambda^2+(p-q)^2)$ is included with the gluon propagator. The
resulting integral is finite $ \forall \Lambda<\infty$, and develops a
logarithmic divergence when the regularisation is removed; i.e.,
$\Lambda\rightarrow\infty$, which is the final stage of any calculation.  We
emphasise that it is only with a translationally invariant regularisation
scheme that Ward-Takahashi identities can be preserved, something that is
crucial to ensuring, e.g., axial-vector current conservation in the chiral
limit.
 
The general form of the dressed-quark propagator is 
\begin{equation} 
S(p) =\frac{Z(p^2;\zeta^2)}{i \gamma \cdot p  +  M(p^2)} \label{Def:MZ} 
\end{equation} 
where the Lorentz scalar functions $Z(p^2;\zeta^2)$ and $M(p^2)$ are
respectively referred to as the quark wave function renormalisation and
running quark mass (or dressed-quark mass-function).  The renormalisation
constants $Z_2$ and $Z_4$ are determined by solving the gap equation, Eq.\
(\ref{quarkdse}), subject to the renormalisation condition that at some large
spacelike $\zeta^2$
\begin{equation} 
\label{renormcond} S^{-1}(p)\big|_{p^2=\zeta^2} =i \gamma \cdot p + m(\zeta)\,, 
\end{equation} 
where $m(\zeta)$ is the renormalised current-quark mass.  The renormalised
dressed-quark propagator is independent of the regularisation mass-scale,
$\Lambda$.  It depends on the renormalisation point, $\zeta$, in a manner
prescribed by the theory's dynamics.  This dependence is expressed in the
calculable $\zeta$-dependence of the wave function renormalisation.  The
dressed-quark mass-function is independent of the regularisation mass-scale
and of the renormalisation point.  Once the renormalisation scheme has been
faithfully applied the regularisation mass-scale may be removed to infinity.
 
Given the form of the dressed gluon propagator, $D_{\mu\nu}(p-q)$, and
dressed quark-gluon vertex, $\Gamma_{\mu}^{i}(q,p)$, it is straightforward to
determine the corresponding dressed quark propagator using well-established
numerical methods.  A particularly important characteristic of the non-linear
integral equation in Eq.\ (\ref{quarkdse}) is that it yields a nonzero
solution for $M(p^2)$ in the chiral limit \cite{mr97,mrt98}:
\begin{equation} 
\label{chirallimit} 
Z_{4}(\zeta,\Lambda) \, m(\zeta) \equiv 0\,,\;  \Lambda \gg \zeta \,,
\end{equation} 
if, and only if, there is sufficient infrared support in the integrand.  This
is dynamical chiral symmetry breaking, which we discuss further in Sec.\
\ref{Sec:Chiral}.  It is noteworthy that for finite $\zeta$ and $\Lambda \to
\infty$, the left hand side (l.h.s.) of Eq.\ (\ref{chirallimit}) is
identically zero, by definition, because the mass term in QCD's Lagrangian
density is renormalisation-point-independent.  The condition specified in
Eq.\ (\ref{chirallimit}), on the other hand, effects the result that at the
(perturbative) renormalisation point there is no mass-scale associated with
explicit chiral symmetry breaking, which is the essence of the chiral limit.

%========================================================= 
% Rainbow approximation 
%========================================================= 
As will now be plain, the kernel of the gap equation, Eq.\ (\ref{quarkdse}),
is formed from the product of the dressed-gluon propagator and
dressed-quark-gluon vertex.  The equation is therefore coupled to the DSEs
satisfied by these functions.  Those equations in turn involve other
$n$-point functions and hence a tractable problem is only realised once a
truncation scheme is specified.  At least one nonperturbative, chiral
symmetry preserving truncation exists \cite{truncscheme,detmold} and the
first term in that scheme is the renormalisation-group-improved rainbow gap
equation, wherein the self-energy, Eq.\ (\ref{quarkSelf}), assumes the form
\begin{equation} 
\int_q^\Lambda \! {\cal G}(Q^2)\, D_{\mu\nu}^{\rm free} (Q) 
\frac{\lambda^a}{2}\gamma_\mu \, S(q) \, \frac{\lambda^a}{2} \gamma_\nu\,, 
\label{DSEAnsatz} 
\end{equation} 
where $Q= p-q$ and $D_{\mu\nu}^{\rm free} (Q)$ is the Landau gauge free-gluon
propagator.

In Eq.\ (\ref{DSEAnsatz}), ${\cal G}(Q^2)$ is an effective interaction, which
expresses the combined effect of dressing both the gluon propagator and
quark-gluon vertex consistent with the constraints imposed by, e.g., vector
and axial-vector Ward-Takahashi identities.  Asymptotic freedom entails
\begin{equation}
{\cal G}(Q^2) = 4\pi\, \alpha(Q^2)\,,\; Q^2\gtrsim 2\, {\rm GeV}^2;
\end{equation}
viz., the effective interaction is proportional to the strong running
coupling in the ultraviolet.  However, its form is unknown for $Q^2 \lesssim
2\,$GeV$^2$.  An explanation of many diverse hadron phenomena has been
obtained by modelling this behaviour \cite{mrrev} but our goal is
different. Hereinafter we explore the ramifications of employing the
dressed-gluon propagator inferred from numerical simulations of lattice-QCD
in building the effective interaction.
 
\section{Analysing lattice data \label{Anallattice}}
\subsection{Lattice gluon propagator \label{sec:latticegluon} } 
In Ref.\ \cite{Leinweber:1998uu} the Landau gauge dressed gluon propagator
was computed using quenched lattice-QCD configurations and the result was
parametrised as:
\begin{equation} 
D(k^2) = Z_{g} \left[ 
  \frac{A \Lambda_{g}^{2\alpha}}{(k^2 + \Lambda_{g}^2)^{1+\alpha}} + 
    \frac{L(k^2,\Lambda_{g})}{k^2+\Lambda_{g}^2} 
           \right], 
\label{ModelGluon} 
\end{equation} 
with 
\begin{equation} 
\begin{array}{lllllll} 
A & = & 9.8^{+0.1}_{-0.9}, & \; & 
\Lambda_{g} &=& 1.020 \pm 0.1 \pm 0.025 \, {\rm GeV},\\[1ex] 
\alpha &=&2.2^{+0.1+0.2}_{-0.2-0.3}, &\;& 
Z_{g}&=&2.01^{+0.04}_{-0.05}, 
\end{array} 
\end{equation} 
where the first pair of errors are statistical and the second, when present,
denote systematic errors associated with finite lattice spacing and volume.
In the simulation the lattice spacing $a=1/[1.885\,$GeV$]$.  The numerator in
the second term of Eq.\ (\ref{ModelGluon}) is
\begin{equation} 
 L(k^2,\Lambda_{g}) = 
     \left( \frac{1}{2} 
       \ln\left[ 
         (k^2+\Lambda_{g}^2)(\frac{1}{k^2}+\frac{1}{\Lambda_{g}^2})\right] 
     \right)^{-d_{D}}, 
     \label{PQCDGluon} 
\end{equation} 
with $d_{D} = [39 - 9 \, \xi - 4 N_{f}]/[2(33 - 2 N_{f})]$, an expression
which ensures the parametrisation expresses the correct one-loop behaviour at
ultraviolet momenta.  In the quenched, Landau-gauge study, $N_f=0$ , $\xi=0$,
so
\begin{equation} 
%d_D= \frac{13}{22}\,. 
d_D= 13/22\,.
\end{equation} 
 
%======================================================== 
\subsection{Effective quark-gluon vertex \label{Sec:EffVertex} } 
%======================================================== 
We have noted that for $Q^2\gtrsim 2\,$GeV$^2$ in Eq.\ (\ref{DSEAnsatz})
\begin{equation} 
\label{QCDG} {\cal G}(Q^2) = \frac{4\pi^2\gamma_m}{\ln( Q^2/\Lambda_{\rm 
QCD}^2)}, 
\end{equation} 
where \mbox{$\gamma_{m}=$} \mbox{$12/(33 - 2 N_{f})$} is the anomalous mass 
dimension.  To proceed we therefore write 
\begin{equation} 
\label{effective} 
\frac{1}{Q^2}\, {\cal G}(Q^2) =  D(Q^2)\, \Gamma_1(Q^2)\,, 
\end{equation} 
with $D(Q^2)$ given in Eq.\ (\ref{ModelGluon}) and 
\begin{equation} 
\label{vQ2} \Gamma_1(Q^2)= 4 \pi^2\gamma_m\, \frac{1}{Z_g}\,\frac{
[\frac{1}{2}\ln(\tau + Q^2/\Lambda_{g}^2)]^{d_D}} {[\ln(\tau +
Q^2/\Lambda_{\rm QCD}^2)]}\, v(Q^2)\,,
\end{equation} 
where $\tau=e^2-1>1$ is an infrared cutoff.  Equation (\ref{effective})
factorises the effective interaction into a contribution from the lattice
dressed-gluon propagator multiplied by a contribution from the vertex, which
we shall subsequently determine phenomenologically.

We remark that the renormalisation-group-improved rainbow truncation retains
only that single element of the dressed-quark-gluon vertex which is
ultraviolet divergent at one-loop level and this explains the simple form of
Eq.\ (\ref{vQ2}).  Systematic analyses of corrections to the rainbow
truncation show $\Gamma_1$ to be the dominant amplitude of the dressed
vertex: the remaining amplitudes do not significantly affect observables
\cite{detmold}.  In proceeding phenomenologically solely with $\Gamma_1$, we
force $v(Q^2)$ to assume the role of the omitted amplitudes to the maximum
extent possible.
 
In Eq.\ (\ref{vQ2}), so long as $v(Q^2) \simeq 1$ for $Q^2\gtrsim
2\,$GeV$^2$, Eq.\ (\ref{QCDG}) is satisfied and consequently the rainbow gap
equation preserves the renormalisation group flow of QCD at one-loop.  We
therefore consider a simple \textit{Ansatz} with this property:
\begin{equation} 
  v(Q^2) = 
       \frac{a_v(m) + Q^2 /\Lambda_{g}^2} 
        {b + Q^2/\Lambda_{g}^2}\,,
    \label{VIR} 
\end{equation} 
where 
\begin{equation} 
a_v(m) = \frac{a_1}{1 + a_2 \, [m(\zeta)/\Lambda_{g}] + a_3 \, 
[m(\zeta)/\Lambda_{g}]^2 } 
\label{VIRParam} 
\end{equation} 
and $a_{1,2,3}$ and $b$ are dimensionless parameters, which are fitted by
requiring that the gap equation yield a solution for the dressed-quark
propagator that agrees well pointwise with the results obtained in numerical
simulations of quenched lattice-QCD \cite{bowman1,bowman2}.  It is important
to note that a good fit to lattice data is impossible unless $a_v(m)$ depends
on the current-quark mass.  While more complicated forms are clearly
possible, the \textit{Ansatz} of Eq.\ (\ref{VIRParam}) is adequate.
 
%========================================================== 
\subsection{Fit to lattice results \label{Sec:LatticeFit}} 
%=========================================================== 
Now, to be explicit, the parameters in Eqs.\ (\ref{VIR}), (\ref{VIRParam})
were determined by the following procedure.  The rainbow gap equation; viz.,
Eq. (\ref{quarkdse}) simplified via Eq.\ (\ref{DSEAnsatz}), was solved using
the effective interaction specified by Eqs.\
(\ref{effective})--(\ref{VIRParam}), with $D(Q^2)$ exactly as given in Eq.\
(\ref{ModelGluon}).
 
The ultraviolet behaviour of the mass function, $M(p^2)$, is determined by
perturbative QCD and is therefore model independent.  Hence the current-quark
mass, $m(\zeta)$, was fixed by requiring agreement between the DSE and
lattice results for $M(p^2)$ on $p^2 \gtrsim 1\,$GeV$^2$.  We selected three
lattice data sets from Ref.\ \cite{bowman2} and, for consistency with Refs.\
\cite{mr97,mt99}, used a renormalisation point $\zeta= 19\,$GeV, which is
well into the perturbative domain.  This gave
\begin{equation} 
\label{amvalues}
\begin{array}{l|lll} 
a\,m_{\rm lattice} & 0.018 & 0.036 & 0.072 \\\hline 
m(\zeta) ({\rm GeV}) & 0.030 & 0.055 & 0.110 
\end{array}\,. 
\end{equation}
 
The dimensionless parameters $a_{1,2,3}$ and $b$ were subsequently determined
in a simultaneous least-squares fit of DSE solutions for $M(p^2)$ at these
current-quark masses to all the lattice data.  This necessarily required the
gap equation to be solved repeatedly.  Nevertheless, the fit required only
hours on a modern workstation, and yielded:
\begin{equation} 
\label{renmasses} 
\begin{array}{c|c|c||c} 
a_1 & a_2 & a_3 & b \\\hline 
1.5 & 7.35 & 63.0 & 0.005 
\end{array}. 
\end{equation} 
These parameters completely determine the ``best-fit effective-interaction''
and hence our lattice-constrained model for the gap equation's kernel.
 
We emphasise that because the comparison is with simulations of quenched
lattice-QCD, we used $N_f=0$ throughout and
\cite{latticevertex1,latticevertex2,latticevertex3,lambdaquenched1}
\begin{equation} 
\Lambda_{\rm qu-QCD}=0.234\,{\rm GeV}. 
\end{equation} 
The strength of the running strong coupling is underestimated in simulations
of quenched lattice-QCD \cite{lambdaquenched2}.  (NB.\ Halving or doubling
$\Lambda_{\rm qu-QCD}$ has no material quantitative impact on the results
reported in Sec.\ \ref{Sec:Results}, nor does it qualitatively affect our
conclusions.)
 
%....................................................... 
\begin{figure} 
\centerline{\resizebox{0.45\textwidth}{!}{\includegraphics{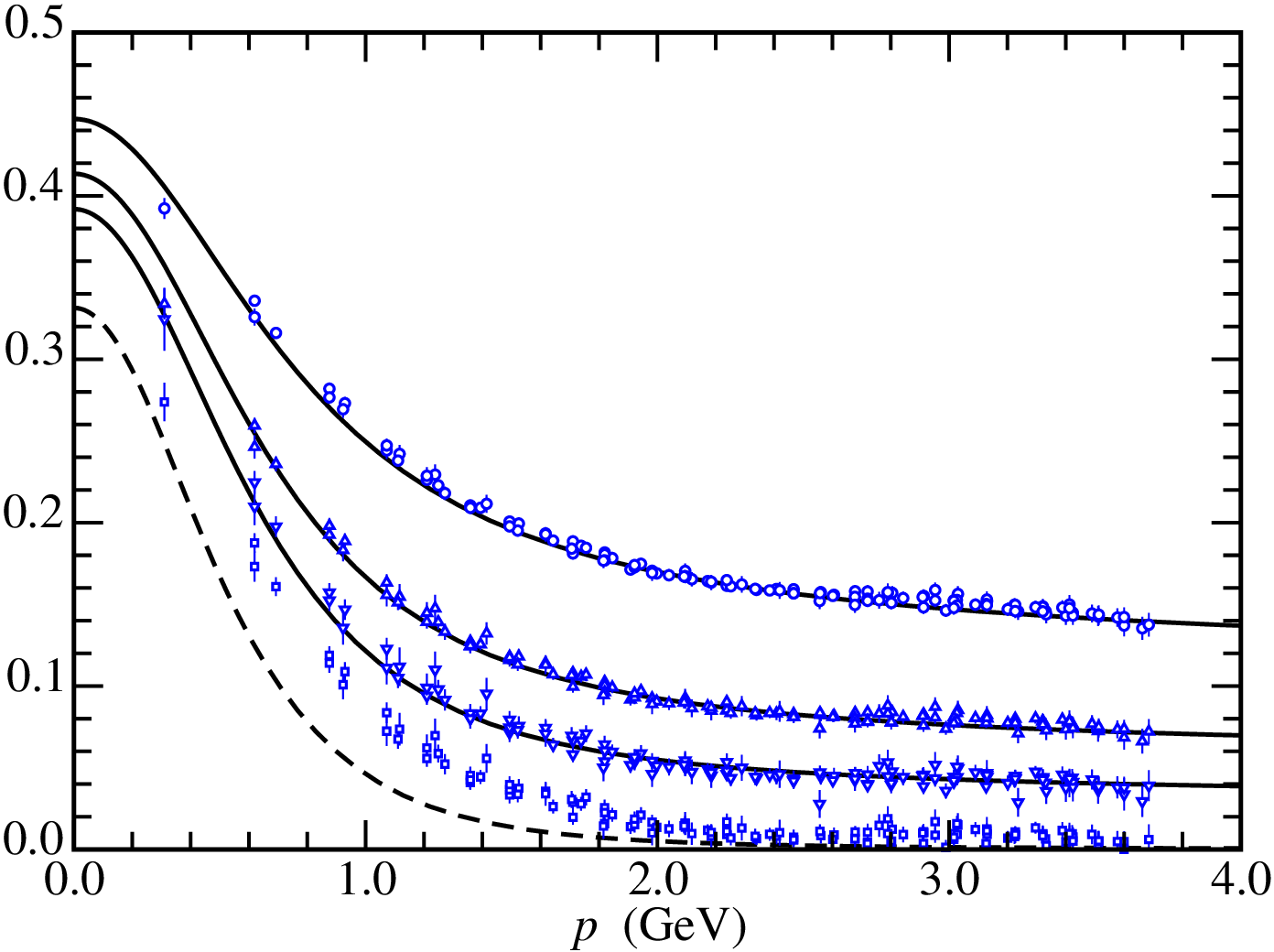}}} 
\caption{\label{Fig:MFit} Data, upper three sets: lattice results for
$M(p^2)$ in GeV at $am$ values in Eq.\ (\protect\ref{amvalues}); lower points
(boxes): linear extrapolation of lattice results \protect\cite{bowman2} to $a
m =0$.  Solid curves: best-fit-interaction gap equation solutions for
$M(p^2)$ obtained using the current-quark masses in Eq.\
(\protect\ref{amvalues}); dashed-curve: gap equation's solution in the chiral
limit, Eq.\ (\protect\ref{chirallimit}). }
\end{figure} 
%....................................................... 
 
\subsubsection{Fidelity and quiddity of the procedure} 
\begin{figure}[t] 
%\vspace*{-5ex}
\centerline{\resizebox{0.45\textwidth}{!}{\includegraphics{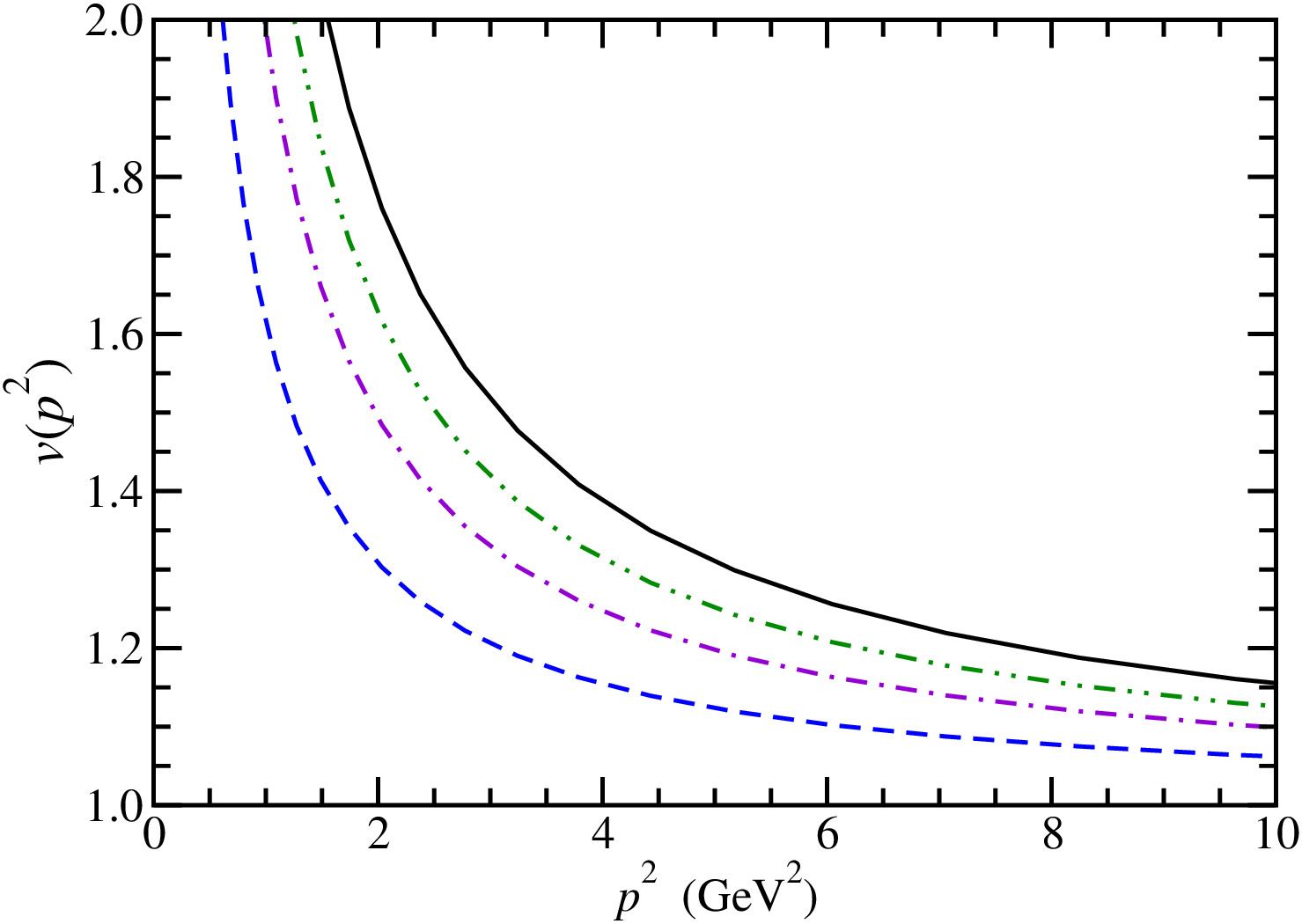}}} 
\caption{\label{Vq2pic} Dimensionless vertex dressing factor: $v(Q^2)$,
defined via Eqs.\ (\protect\ref{VIR}), (\protect\ref{VIRParam}),
(\protect\ref{renmasses}), obtained in the chiral limit (\textit{solid
curve}) and with the current-quark masses in Eq.\ (\protect\ref{amvalues}).
$v(Q^2)$ is finite at $Q^2=0$ and decreases with increasing $m(\zeta)$.}
\end{figure}

In Fig.\ \ref{Fig:MFit} we compare DSE solutions for $M(p^2)$, obtained using
the optimised effective interaction, with lattice results.  In addition, we
depict the DSE solution for $M(p^2)$ calculated in the chiral limit along
with the linear extrapolation of the lattice data to $ a m =0$, as described
in Ref.\ \cite{bowman1}.
It is apparent that the lattice-gluon and lattice-quark propagators can be
correlated via the renormalisation-group-improved gap equation.  That was
achieved via $v(Q^2)$ in Eq.\ (\ref{VIR}), and the required form is depicted
in Fig.\ \ref{Vq2pic}.  Plainly, consistency between the propagators via this
gap equation requires an infrared enhancement of the vertex, as anticipated
in Refs.\ \cite{fred1,hawes,bloch,raya}.
Our inferred form is in semi-quantitative agreement with the result of recent, 
exploratory lattice-QCD simulations of the dressed-quark-gluon vertex 
\cite{latticevertex1,latticevertex2,latticevertex3}. 
 
Dynamical chiral symmetry breaking is another important feature evident in
Fig.\ \ref{Fig:MFit}; viz., the existence of a $M(p^2)\neq 0$ solution of the
gap equation in the chiral limit.  We deduce that DCSB is manifest in
quenched-QCD and, in the following, quantify the magnitude of that effect.
It should be observed that a linear extrapolation to $am=0$ of the lattice
data obtained with nonzero current-quark masses overestimates the mass
function calculated directly as the solution of the gap equation.
 
%....................................................... 
%....................................................... 
 
\begin{figure}[t]
\centerline{\resizebox{0.45\textwidth}{!}{\includegraphics{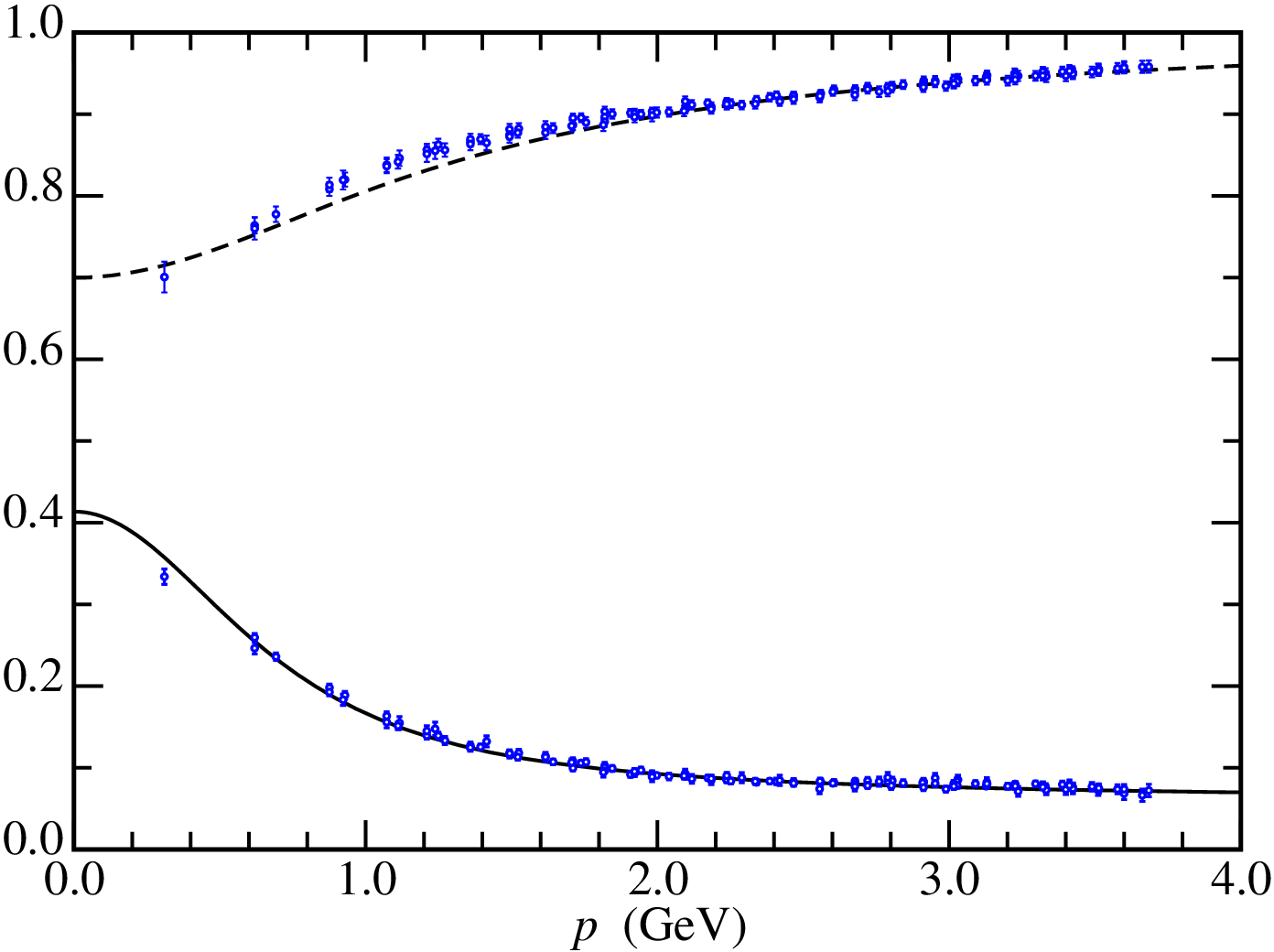}}} 
\caption{Data, quenched lattice-QCD results for $M(p^2)$ and $Z(p^2)$
obtained with $am = 0.036$ \cite{bowman2}; dashed curve, $Z(p^2)$, and solid
curve, $M(p^2)$ calculated from the gap equation with our optimised effective
interaction and $m(\zeta)=55\,$MeV.  (NB.\ $Z(p^2)$ is dimensionless and
$M(p^2)$ is measured in GeV.) \label{Fig:MZFit}}
\end{figure}

Figure \ref{Fig:MZFit} focuses on the lattice simulations for the
intermediate value of the current-quark mass; namely, $am = 0.036$, and
compares lattice output for both $M(p^2)$ and the quark wave function
renormalisation, $Z(p^2)$, with our results.  We emphasise that the form of
$Z(p^2)$ was not used in fitting $v(Q^2)$.  Hence the pointwise agreement
between the gap equation's solution and the lattice result indicates that our
simple expression for the effective interaction captures the dominant
dynamical content and, in particular, that omitting the subdominant
amplitudes in the dressed-quark-gluon vertex is not a serious flaw in this
study.

\subsection{Spectral properties}
\label{sec:confinement}
In a quantum field theory defined by a Euclidean measure \cite{gj81} the
Osterwalder-Schrader axioms \cite{OSA,OSB} are five conditions which any
moment of this measure ($n$-point Schwinger function) must satisfy if it is
to have an analytic continuation to Minkowski space and hence an association
with observable quantities.  One of these is ``OS3'': the axiom of
\textit{reflection positivity}, which is violated if the Schwinger function's
Fourier transform to configuration space is not positive definite.  The space
of observable asymptotic states is spanned by eigenvectors of the theory's
infrared Hamiltonian and no Schwinger function that breaches OS3 has a
correspondent in this space.  Consequently, the violation of OS3 is a
sufficient condition for confinement.  

This connection has long been of interest \protect\cite{cornwall,mn83}, and
is discussed at length in Refs.\ \protect\cite{gastao,stingl1,stingl2}, and
reviewed in Sec.\ 6.2 of Ref.\ \protect\cite{cdragw}, Sec.\ 2.2 of Ref.\
\protect\cite{bastirev} and Sec.\ 2.4 of Ref.\ \protect\cite{reinhardrev}.
It suggests and admits a practical test \cite{fred1} that has been exploited
in Refs.\
\cite{FischerAlkofer03,fred1,hawes,axelT,axelmu,axelreinhard,pieterqed3,gomezdumm,sauli,mandarconf}
and which for the quark $2$-point function is based on the behaviour of
\begin{eqnarray}
\Delta_S(T) & = & \int \! d^3 x\, \int\! \frac{d^4 p}{(2\pi)^4}\,{\rm e}^{i
 p\cdot x}\, \sigma_S(p^2)\\
& =& \frac{1}{\pi} \int_0^\infty\! d\varepsilon\, \cos(\varepsilon \, T)
\,\sigma_S(\varepsilon^2)\,,
\end{eqnarray}
where $\sigma_S$ is the Dirac-scalar projection of the dressed-quark
propagator.  For a noninteracting fermion with mass $\mu$,
\begin{equation}
\label{freeD}
\Delta_S^{\rm free}(T) = \frac{1}{\pi} \int_0^\infty\! d\varepsilon \
\cos(\varepsilon \, T) \, \frac{\mu}{\varepsilon^2+\mu^2}= \frac{1}{2}\, {\rm
e}^{-\mu T}.
\end{equation}
The r.h.s.\ is positive definite.  It is also plainly related via analytic
continuation ($T \to i t$) to the free-particle solution of the Minkowski
space Dirac equation.  The existence of an associated asymptotic state is
indubitable.

%....................................................... 
%....................................................... 

If instead one encountered a theory in which
\cite{stingl1,stingl2,mandarconf}
% \begin{equation}
% \sigma_S(p^2) = \frac{\mu \, (p^2+\mu^2)}{(p^2+\mu^2)^2 + \rho^4},
% \end{equation}
\begin{equation}
\label{stinglS}
\sigma_S(p^2) = \frac{\mu}{2} \left[ \frac{1}{p^2 + \mu^2 - i \rho^2} +
\frac{1}{p^2 + \mu^2 + i \rho^2} \right],
\end{equation}
a function with poles at $p^2+ \sigma^2 \exp(\pm i \theta)=0$, where
\begin{equation}
\sigma^4 = \mu^4+\rho^4\,,\; \tan\theta=\rho^2/\mu^2,
\end{equation}
then
\begin{equation}
\label{stinglD}
\Delta_S(T) = \frac{\mu}{2 \sigma}\, {\rm e}^{-\sigma T
\cos\frac{\theta}{2}}\, \cos\left( \sigma T \sin\frac{\theta}{2} +
\frac{\theta}{2}\right) \,.
\end{equation}
This Fourier transform has infinitely many, regularly spaced zeros and hence
OS3 is violated.  Thus the fermion described by this Schwinger function has
no correspondent in the space of observable asymptotic states.

It is readily apparent that Eq.\ (\ref{stinglS}) evolves to a free particle
propagator when $\rho\to 0$.  This limit is expressed in Eq.\ (\ref{stinglD})
via $\theta\to 0$, $\sigma\to\mu$, wherewith Eq.\ (\ref{freeD}) is recovered;
a result that is tied to the feature that the first zero of $\Delta_S(T)$ in
Eq.\ (\ref{stinglD}) occurs at
\begin{equation}
z_1= \frac{\pi - \theta}{2\, \sigma} \,\csc\frac{\theta}{2}
\end{equation}
and hence $z_1 \to \infty$ for $\rho \to 0$ \cite{axelT,axelmu}.  

%....................................................... 
\begin{figure}[t]
\centerline{\resizebox{0.45\textwidth}{!}{%
\includegraphics{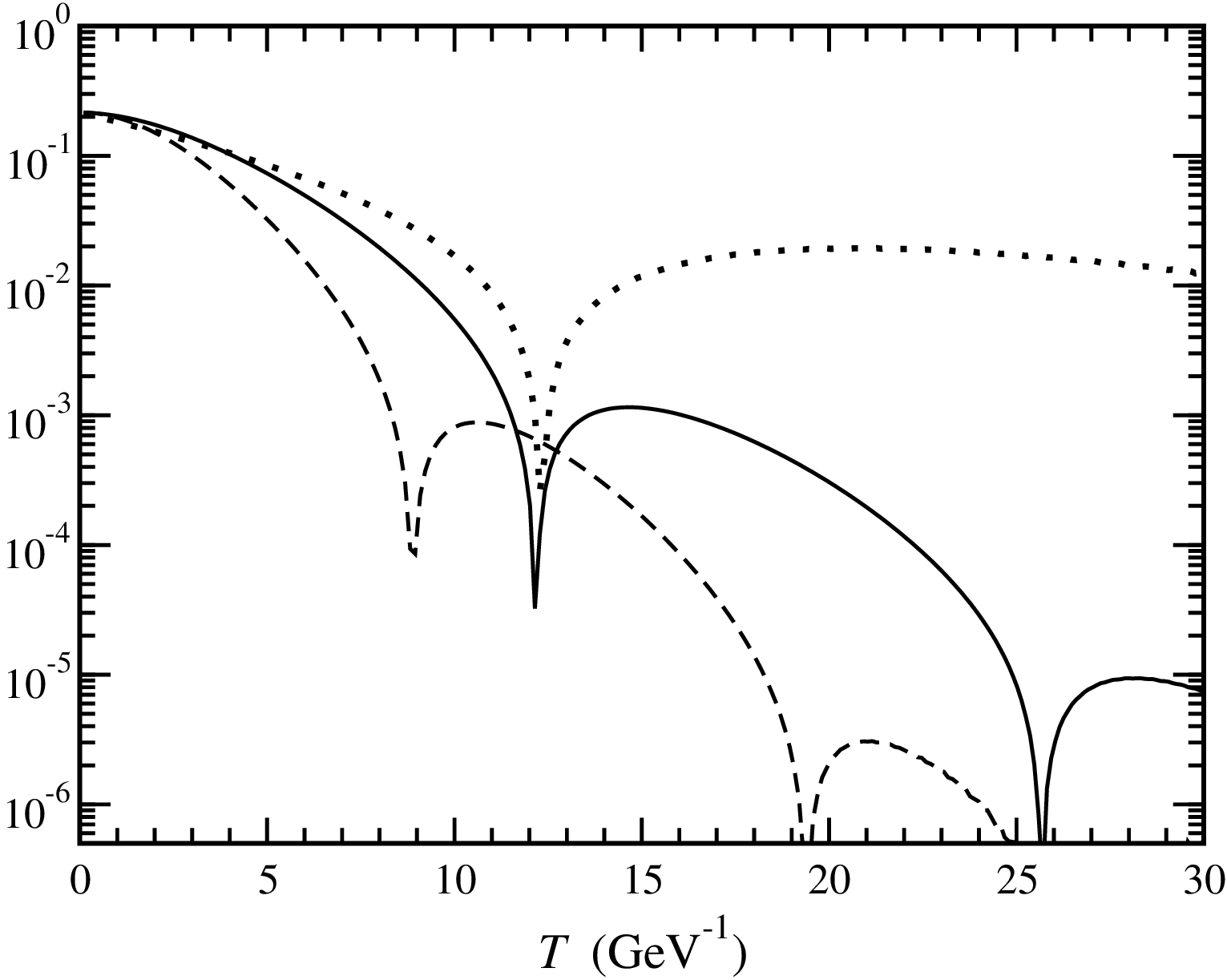}}}
\caption{$|\Delta_S(T)|$ obtained from: the chiral limit gap equation
solution calculated using our lattice-constrained kernel, solid curve; Eq.\
(\protect\ref{stinglS}) with $\sigma=0.13\,$GeV, $\theta=\pi/2.46$, dotted
curve; the model of Ref.\ \protect\cite{mt99}, dashed curve.
\label{fig:conf}}
\end{figure} 
%....................................................... 

We calculated $\Delta_S(T)$ using the gap equation solutions discussed above
and the form of $|\Delta_S(T)|$ obtained with the chiral limit solution is
depicted in Fig.\ \ref{fig:conf}.  The violation of OS3 is manifest in the
appearance of cusps: $\ln|\Delta_S(T)|$ is negative and infinite in magnitude
at zeros of $\Delta_S(T)$.  The differences evident in a comparison with the
result obtained from Eq.\ (\ref{stinglD}) indicate that the singularity
structure of the dressed-quark $2$-point Schwinger function obtained from the
lattice-constrained kernel is more complicated than just a single pair of
complex conjugate poles.  However, the similarities suggest that this picture
may serve well as an idealisation \cite{mandarconf}.  (NB.\ Qualitatively
identical results are obtained using $\sigma_V(p^2)$, the Dirac-vector
projection of the dressed-quark propagator, instead of $\sigma_S(p^2)$.)

Figure \ref{fig:conf} also portrays the result obtained with the effective
interaction proposed in Ref.\ \cite{mt99} and used efficaciously in studies
of meson properties \cite{mrrev}.  Significantly, the first zero appears at a
smaller value of $T$ in this case.  Using the model of Eq.\ (\ref{stinglS})
as a guide, that shift indicates a larger value of $\sigma$.  This sits well
with the fact that the mass-scale generated dynamically by the interaction of
Ref.\ \cite{mt99} is larger than that produced by the interaction used
herein, which is only required to correlate quenched lattice data for the
gluon and quark two-point functions.

From the results and analysis reported in this subsection, we deduce that
light-quarks do not appear in the space of observable asymptotic states
associated with quenched-QCD; an outcome anticipated in Ref.\ \cite{raya}.

%------------------------------------------------------------------------ 
\section{Chiral and Physical Pion Observables \label{Sec:Results}} 
%===================================================================== 
\subsection{Chiral limit \label{Sec:Chiral}} 
%===================================================================== 
The scale of DCSB is measured by the value of the renormalisation-point 
dependent vacuum quark condensate, which is obtained directly from the chiral 
limit dressed-quark propagator \cite{mrt98,mr97}: 
\begin{equation} 
\label{qbq0} - \langle \bar{q}{q} \rangle_{\zeta}^0 = 
\lim_{\Lambda\rightarrow\infty} 
 Z_{4}(\zeta,\Lambda) \; N_c \; {\rm tr} \!\int_{q}^{\Lambda}\!\!\! S_{0}(q),
\end{equation} 
where ``tr'' denotes a trace only over Dirac indices and the index ``$0$''
labels a quantity calculated in the chiral limit, Eq.\ (\ref{chirallimit}).
In Eq.\ (\ref{qbq0}) the gauge parameter dependence of the renormalisation
constant $Z_4$ is precisely that required to ensure the vacuum quark
condensate is gauge independent.  This constant is fixed by the
renormalisation condition, Eq.\ (\ref{renormcond}), which entails
\begin{equation} 
Z_4(\zeta,\Lambda) = - \frac{1}{m(\zeta)}\, \frac{1}{4}\, {\rm tr} \,
\Sigma^\prime(\zeta,\Lambda) 
\end{equation} 
wherein the right hand side is well-defined in the chiral limit \cite{mr97}.
(It may also be determined by studying the fully amputated
pseudoscalar-quark-antiquark $3$-point function \cite{mrt98}.)  A
straightforward calculation using our chiral limit result; i.e, the
propagator corresponding to the dashed-curve in Fig.\ \ref{Fig:MFit}, yields
\begin{equation} 
  -\langle \bar{q}q\rangle^0_{\zeta = 19~{\rm GeV}} =(0.22\,{\rm GeV})^{3}. 
   \label{Cond19} 
\end{equation} 
 
To evolve the condensate to a ``typical hadronic scale,'' e.g.,
$\zeta=1\,$GeV, one may use \cite{langfeld}
\begin{equation} 
\langle \bar{q}q\rangle^0_{\zeta^\prime} = Z_4(\zeta^\prime,\zeta) 
Z_2^{-1}(\zeta^\prime,\zeta)\, \langle \bar{q}q\rangle^0_{\zeta} =: 
Z_m(\zeta^\prime,\zeta) \langle \bar{q}q\rangle^0_{\zeta}, 
\end{equation} 
where $Z_m$ is the gauge invariant mass renormalisation constant.
Contemporary phenomenological approaches employ the one-loop expression for
$Z_m$ and following this expedient we obtain, practically as a matter of
definition,
\begin{eqnarray} 
\nonumber
-\langle \bar{q}q\rangle^0_{1\,{\rm GeV}} &=& \left( \frac{\ln[1/\Lambda_{\rm 
qu-QCD}]}{\ln[19/\Lambda_{\rm qu-QCD}]}\right)^{\gamma_m} \!\!\! (-\langle 
\bar{q}q\rangle^0_{19\,{\rm GeV}})\\ 
&=&  (0.19\,{\rm GeV})^2, 
\end{eqnarray} 
which may be compared with a best-fit phenomenological value \cite{derek}:
$(0.24 \pm 0.01\,{\rm GeV})^3$.  It is notable that DSE models which
efficaciously describe light-meson physics; e.g., Refs.\ \cite{mr97,mt99},
give $\langle \bar{q}q\rangle^0_{1\,{\rm GeV}} = -(0.24\,{\rm GeV})^3$.

Our gap equation assisted estimate therefore indicates that the chiral
condensate in quenched-QCD is a factor of two smaller than that which is
obtained from analyses of strong interaction observables.  These results are
in quantitative agreement with Ref.\ \cite{raya}.

A fit to the linear extrapolation of the lattice data; viz., to the boxes in
Fig.\ \ref{Fig:MFit}, gives a significantly larger value \cite{bowman2}:
$-\langle\bar{q}q\rangle^0_{1\,{\rm GeV}}= (0.270\pm 0.027\,{\rm GeV})^3$.
However, the error is purely statistical.  The systematic error, to which the
linear extrapolation must contribute, was not estimated.  That may be
important given the discrepancy, conspicuous in Fig.\ \ref{Fig:MFit}, between
our direct evaluation of the chiral limit mass function and the linear
extrapolation of the lattice data to $am=0$: the linear extrapolation lies
well above the result of our chiral limit calculation.
 
%....................................................... 
\begin{figure}[t] 
\centerline{\resizebox{0.45\textwidth}{!}{\includegraphics{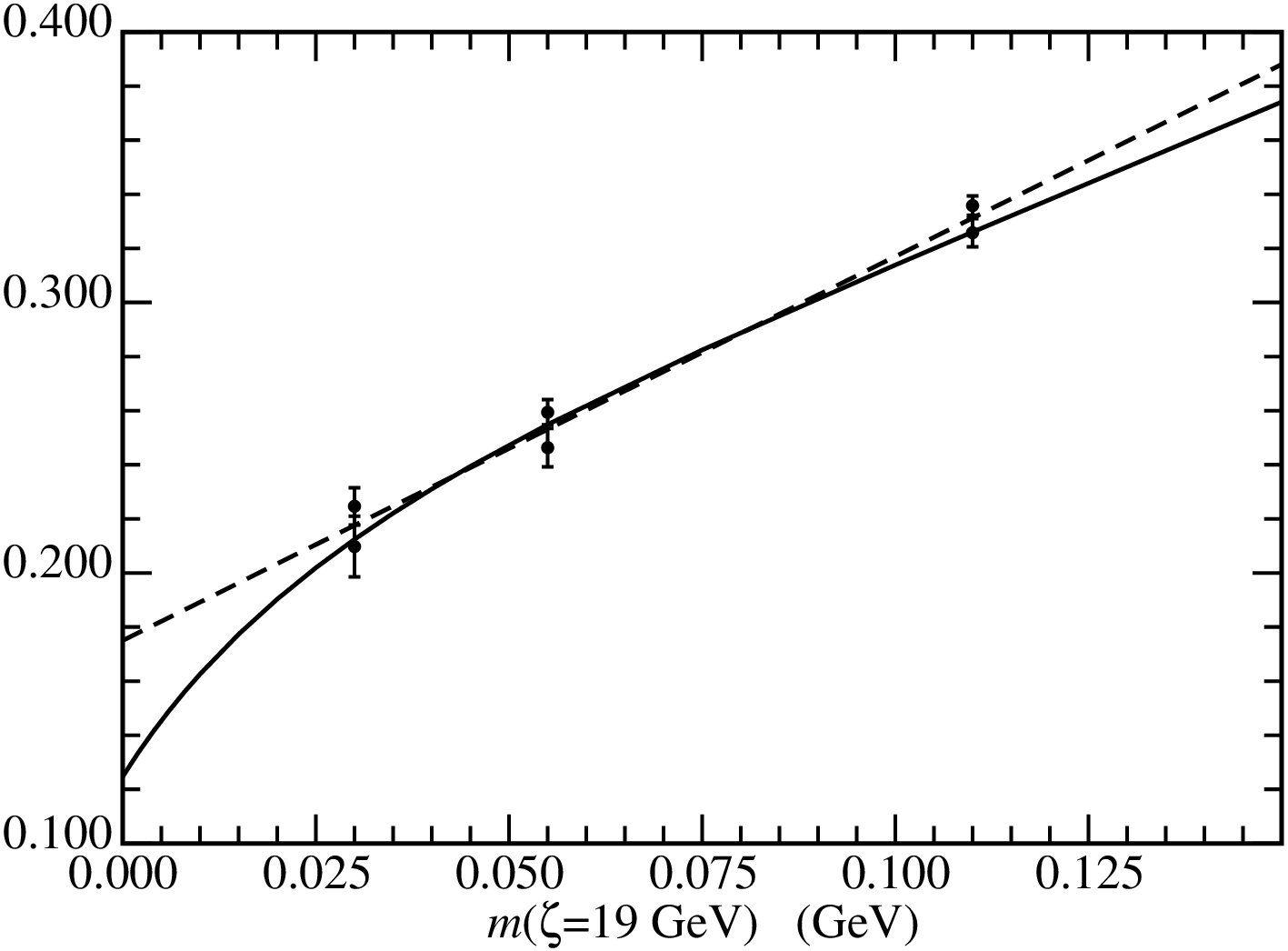}}} 
\caption{$M(p_{\rm IR}^2=0.38\,{\rm GeV}^2)$, in GeV, as a function of the
current-quark mass.  Solid curve, our result; circles, lattice data for $am$
in Eq.\ (\protect{amvalues}) \protect\cite{bowman2}; dashed-line, linear fit
to the lattice data, Eq.\ (\protect\ref{M0linear}).
\label{Fig:Mvsm} } 
\end{figure} 
%....................................................... 
 
This point may be illustrated further.  In Fig.\ \ref{Fig:Mvsm} we plot
$M(p^2=p_{\rm IR}^2)$, where $p_{\rm IR} = 0.62\,{\rm GeV}$, as a function of
the current-quark mass (\textit{solid curve}).  This value of the argument
was chosen because it is the smallest $p^2$ for which there are two lattice
results for $M(p^2)$ at each current-quark mass in Eq.\ (\ref{amvalues}).
Those results are also plotted in the figure.  It is evident that on the
domain of current-quark masses directly accessible in lattice simulations,
the lattice and DSE results lie on the same linear trajectory, of which
\begin{equation}
\label{M0linear}
M(p_{\rm IR}^2=0.38\,{\rm GeV}^2,m(\zeta))= 0.18 + 1.42\,m(\zeta)
\end{equation}
provides an adequate interpolation.  However, as apparent in the figure, this
fit on $am \in [ 0.018,0.072]$ provides a poor extrapolation to $m(\zeta) =
0$, giving a result $40$\% too large.

\begin{figure}[t] 
\centerline{\resizebox{0.45\textwidth}{!}{\includegraphics{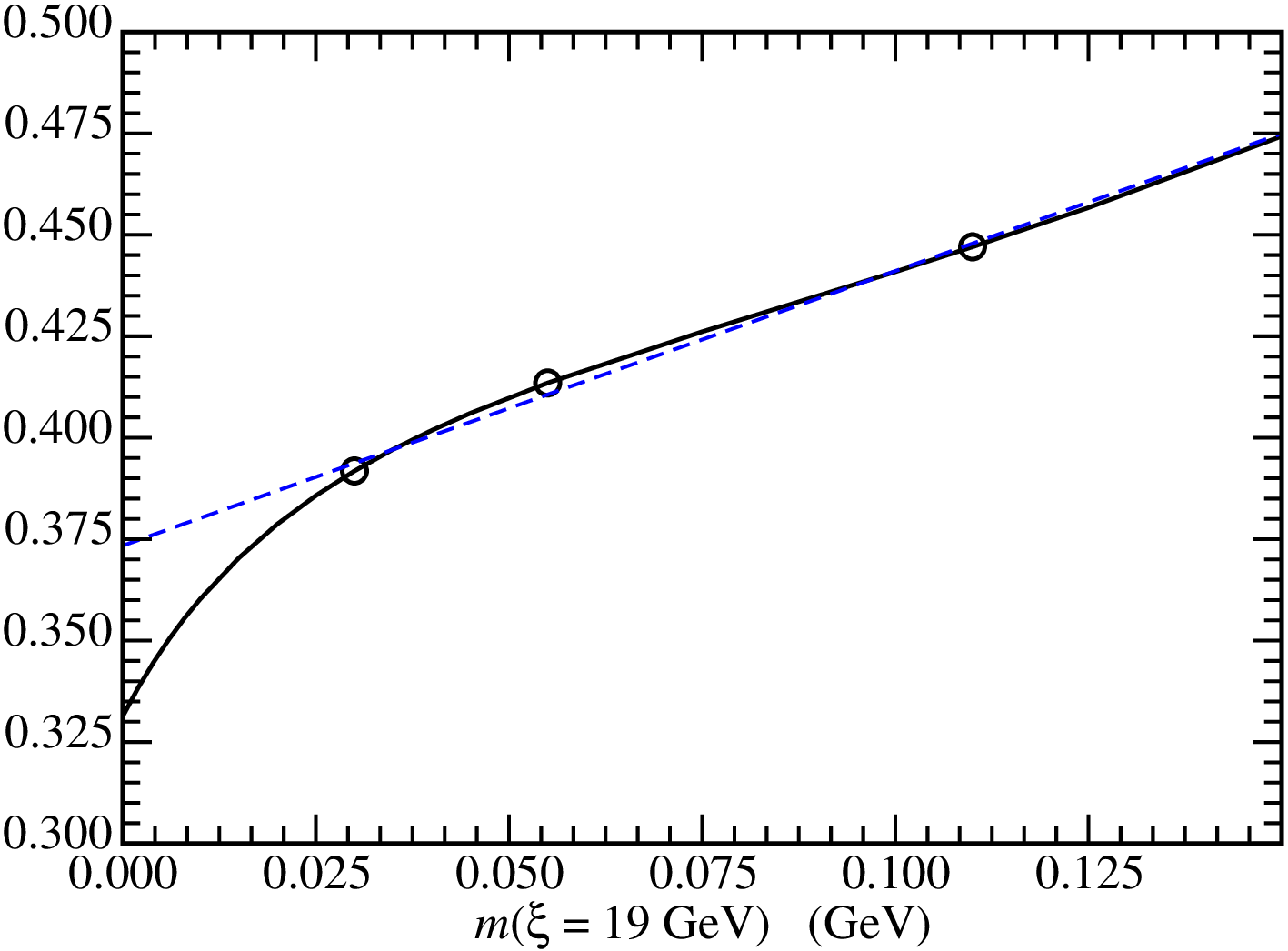}}} 
\caption{Solid curve, calculated $M(p^2=0)$, in GeV, as a function of the
current-quark mass $m(\zeta)$.  The circles mark the current-quark masses in
Eq.\ (\protect\ref{amvalues}).  Dashed-line, linear interpolation of our
result for $M(p^2=0)$ on this mass domain.
\label{Fig:fig5} } 
\end{figure}

In Fig.\ \ref{Fig:fig5} we repeat this procedure, focusing solely on our
value of $M(p^2=0$) because directly calculated lattice data are unavailable
at this extreme infrared point and published estimates obtained by
extrapolating functions fitted to the lattice-$p^2$-dependence are
inconsistent \cite{bowman1}.  The pattern observed in Fig.\ \ref{Fig:Mvsm} is
again visible.  On the domain of current-quark masses for which lattice data
are available, the mass-dependence of $M(p^2)$ is well-approximated by a
straight line; namely,
\begin{equation}
\label{M00linear}
M(p^2=0,m(\zeta))= 0.37 + 0.68\,m(\zeta)\,,
\end{equation}
but the value of $M(0)$ determined via extrapolation to $m(\zeta)=0$ is
$14$\% too large.

In all cases our calculated result possesses significant curvature.  At fixed
$p^2$, $M(p^2,m(\zeta))$ is a monotonically increasing function of $m(\zeta)$
but, while it is concave-down for $m(\zeta)\lesssim 0.1\,$GeV, it inflects
thereafter to become concave-up.  In addition, at fixed $m(\zeta)$, $M(p^2)$
is a monotonically decreasing function of $p^2$.  It follows that a linear
extrapolation determined by data on $am \in[0.018,0.072]$ will necessarily
overestimate $M(p^2)$ at all positive values of $p^2$.  The figures
illustrate that the error owing to extrapolation increases with increasing
$p^2$ and hence a significant overestimate can be anticipated on the domain
in which the condensate was inferred from lattice data.
 
%....................................................... 
%....................................................... 
 
It is straightforward to understand the behaviour evident in Figs.\
\ref{Fig:Mvsm},  \ref{Fig:fig5} qualitatively.  The existence of DCSB means
that in the neighbourhood of the chiral limit a mass-scale other than the
current-quark mass determines the magnitude of $M(p^2)$.  As the
current-quark mass increases from zero, its magnitude will come to affect
that of the mass function. The gap equation is a nonlinear integral equation
and hence this evolution of the mass-dependence of $M(p^2)$ will in general
be nonlinear.  Only at very large values of the current-quark mass will this
scale dominate the behaviour of the mass function, as seen in studies of
heavy-quark systems \cite{misha}, and the evolution become linear.

\subsection{Pion properties}
The renormalised homogeneous Bethe-Salpeter equation (BSE) for the
isovector-pseudoscalar channel; i.e., the pion, is
\begin{equation} 
\label{genbsepi} 
\left[\Gamma_\pi^j(k;P)\right]_{tu} =  \int^\Lambda_q 
\,[\chi_\pi^j(q;P)]_{sr} \,K^{rs}_{tu}(q,k;P)\,, 
\end{equation} 
where $k$ is the relative momentum of the quark-antiquark pair, $P$ is their
total momentum; and
\begin{equation} 
\label{chiM} 
\chi_\pi^j(q;P)=S(q_+) \Gamma_\pi^j(q;P) S(q_-)\,, 
\end{equation} 
with $\Gamma_\pi^j(k;P)$ the pion's Bethe-Salpeter amplitude, which has the
general form
\begin{eqnarray} 
\nonumber 
\lefteqn{\Gamma_\pi^j(k;P)
 = \tau^j \gamma_5 \left[ i E_\pi(k;P) +
\gamma\cdot P F_\pi(k;P) \rule{0mm}{5mm}\right.}\\
& + & \left. \rule{0mm}{5mm} \gamma\cdot k \,k \cdot P\, G_\pi(k;P) + 
\sigma_{\mu\nu}\,k_\mu P_\nu \,H_\pi(k;P) \right] \!.  \label{genpibsa} 
\end{eqnarray}
In Eq.\ (\ref{genbsepi}), $K(q,k;P)$ is the fully-amputated quark-antiquark
scattering kernel, and the axial-vector Ward-Takahashi identity requires that
this kernel and that of the gap equation must be intimately related.  The
consequences of this are elucidated in Refs.\ \cite{truncscheme,detmold} and
in the present case they entail
\begin{eqnarray}
\nonumber \lefteqn{K_{tu}^{rs}(q,k;P) = - {\cal G}((k-q)^2)  }\\
&&  \times \, D_{\mu\nu}^{\rm free} (q-k) \,
\left[\rule{0mm}{0.7\baselineskip} 
        \frac{\lambda^a}{2}\gamma_\mu \right]_{ts} 
\left[\rule{0mm}{0.7\baselineskip} 
        \frac{\lambda^a}{2}\gamma_\nu\right]_{r u},
\end{eqnarray}
which provides the renormalisation-group-improved lad\-der-truncation of the
BSE.  The efficacy of combining the renormalisation-group-improved
rainbow-DSE and ladder-BSE truncations is exhibited in Ref.\ \cite{mrrev}.
In particular, it guarantees that in the chiral limit the pion is both a
Goldstone mode and a bound state of a strongly dressed-quark and -antiquark,
and ensures consistency with chiral low energy theorems
\cite{bicudo1,bicudo2}.
 
All the elements involved in building the kernel of the pion's BSE were
determined in the last section and hence one can solve for the pion's mass
and Bethe-Salpeter amplitude.  To complete this exercise practically we
consider 
\begin{equation}
\label{genbsepilambda} 
[1-\ell(P^2)]\, \left[\Gamma_5^j(k;P)\right]_{tu} = \int^\Lambda_q
\,[\chi_5^j(q;P)]_{sr} \,K^{rs}_{tu}(q,k;P)\,.
\end{equation}
This equation has a solution for arbitrary $P^2$ and solving it one obtains a
trajectory $\ell(P^2)$ whose first zero coincides with the bound state's
mass, at which point $\Gamma_5^j(k;P)$ is the true bound state amplitude.  In
general, solving for $\ell(P^2)$ in the physical domain: $P^2<0$, requires
that the integrand be evaluated at complex values of its argument.  However,
as $m_\pi^2\ll M^2(0)$; i.e., the magnitude of the zero is much smaller than
the characteristic dynamically-generated scale in this problem, we avoid
complex arguments by adopting the simple expedient of calculating
$\ell(P^2>0)$ and extrapolating to locate its timelike zero.  The
Bethe-Salpeter amplitude is identified with the $P^2=0$ solution.  Naturally,
this expedient yields the exact solution in the chiral limit; and in cases
where a comparison with the exact solution has been made for realistic,
nonzero light-quark masses, the error is negligible \cite{pmprivate}, as we
shall subsequently illustrate.

Once its mass and bound state amplitude are known, it is straightforward to
calculate the pion's leptonic decay constant \cite{mrt98}:
\begin{equation} 
\label{fpiexact} f_\pi \,\delta^{ij} \,  2\, P_\mu 
= Z_2\,{\rm tr} \int^\Lambda_q \! \tau^i \gamma_5\gamma_\mu
\chi_\pi^j(q;P)\,. 
\end{equation} 
In this expression, the factor of $Z_2$ is crucial: it ensures the result is
gauge invariant, and cutoff and renormalisation-point independent.  (The
Bethe-Salpeter amplitude is normalised canonically \cite{llewellyn}.)

\begin{table}[t] %[H] add [H] placement to break table across pages 
\begin{ruledtabular} 
\begin{tabular}{c|ll} 
          & Calc. (quenched) & Experiment \\ \hline 
$m_{\pi}$ & 0.1385 & 0.1385 \\ 
$f_{\pi}$ & 0.066  & 0.0924 \\ 
$f_{\pi}^{0}$ & 0.063 &  \\ 
% \hline 
% ${-\langle \bar{q}q\rangle^0_{\zeta=1\;\;}}$ & (0.189)$^3$ &
% (0.236$\pm$0.008)$^3$ \\  
\end{tabular} 
\end{ruledtabular} 
\caption{\label{Table:Results} Pion-related observables calculated using our
lattice-constrained effective interaction.  $m(\zeta=19\,{\rm GeV})=3.3\,$MeV
was chosen to give $m_\pi= 0.1395\,$GeV.  The index ``$0$'' indicates a
quantity obtained in the chiral limit.  }
\end{table} 
%....................................................... 
 
Table \ref{Table:Results} lists values of pion observables calculated using
the effective interaction obtained in Sec.\ \ref{Sec:LatticeFit} by fitting
the quenched-QCD lattice data.  (The nonzero current-quark mass in the table
corresponds to a one-loop evolved value of $m(1\,{\rm GeV})= 5.0\,$MeV.)  To
aid with the consideration of these results we note that unquenched
chiral-limit DSE calculations that accurately describe hadron observables
give \cite{mt99} $f_{\pi}^{0} = 0.090\,$GeV.  We infer from these results
that the pion decay constant in quenched lattice-QCD is underestimated by
$30$\%.  Quantitatively equivalent results were found in Ref.\ \cite{raya}.

The rainbow-ladder truncation of the gap and Bethe-Salpeter equations is
chiral symmetry preserving: without fine-tuning it properly represents the
consequences of chiral symmetry and its dynamical breaking.  The truncation
expresses the model-independent mass formula for flavour-nonsinglet
pseudoscalar mesons \cite{mrt98} a corollary of which, at small current-quark
masses, is the Gell--Mann--Oakes--Renner relation:
\begin{equation}
\label{gmorsimple}
(f_{\pi}^0)^2\, m_{\pi}^{2} = - 2 \,m(\zeta)\,
\langle\bar{q}q\rangle^0_{\zeta} + O(m^2(\zeta))\,.
\end{equation}
Inserting our calculated values on the l.h.s.\ and r.h.s.\ of Eq.\
(\ref{gmorsimple}), we have
\begin{equation}
(0.093\,{\rm GeV})^4 \; {\rm cf.} \; (0.091\,{\rm GeV})^4;
\end{equation}
viz, the same accuracy seen in exemplary coupled DSE-BSE calculations; e.g.,
Ref.\ \cite{mr97}.  This establishes the fidelity of our expedient for
solving the BSE.
 
%------------------------------------------------------------------------ 
\section{Summary \label{Sec:Conclusion}}  
We studied quenched-QCD using a rainbow-ladder truncation of the
Dyson-Schwinger equations (DSEs) and demonstrated that existing results from
lattice simulations of quenched-QCD for the dressed-gluon and \mbox{-quark}
Schwinger functions can be correlated via a gap equation that employs a
renormalisation-group-improved model interaction.
As usual, the ultraviolet behaviour of this effective interaction is fully
determined by perturbative QCD.
For the infrared behaviour we employed an {\it Ansatz} whose parameters were
fixed in a least squares fit of the gap equation's solutions to lattice data
on the dressed-quark mass function, $M(p^2)$, at available current-quark
masses.  With our best-fit parameters the mass functions calculated from the
gap equation were indistinguishable from the lattice results.
The gap equation simultaneously yields the dressed-quark renormalisation
function, $Z(p^2)$, and, without tuning, our results agreed with those
obtained in the lattice simulations.

To correlate the lattice's dressed-gluon and -quark Schwinger functions it
was necessary for the gap equation's kernel to exhibit infrared enhancement
over and above that observed in the gluon function alone.  We attributed that
to an infrared enhancement of the dressed-quark-gluon vertex.  The magnitude
of the vertex modification necessary to achieve the correlation is
semi-quantitatively consistent with that observed in quenched lattice-QCD
estimates of this three-point function.

With a well-defined effective interaction, the gap equation provides a
solution for the dressed-quark Schwinger function at arbitrarily small
current-quark masses and, in particular, in the chiral limit: no
extrapolation is involved.  A kernel that accurately describes dressed-quark
lattice data at small current-quark masses may therefore be used as a tool
with which to estimate the chiral limit behaviour of the lattice Schwinger
function.  Our view is that this method is a more reliable predictor than a
linear extrapolation of lattice data to the chiral limit.  Even failing to
accept this perspective, the material difference between results obtained via
the lattice-constrained gap equation and those found by linear extrapolation
of the lattice data must be cause for concern in employing the latter.

In addition, from a well-defined gap equation it is straightforward to
construct symmetry-preserving Bethe-Salpeter equations whose bound state
solutions describe mesons.  We illustrated this via the pion, and calculated
its mass and decay constants in our DSE model of the quenched theory.

Finally, assuming that existing lattice-QCD data are not afflicted by large
systematic errors associated with finite volume or lattice spacing, we infer
from our analysis that quenched QCD exhibits dynamical chiral symmetry
breaking and dressed-quark two-point functions that violate reflection
positivity but that chiral and physical pion observables are significantly
smaller in the quenched theory than in full QCD.

%------------------------------------------------------------------------ 
\begin{acknowledgments} 
We thank P.O.\ Bowman for the lattice data, and acknowledge valuable
interactions with P.\ Maris, R.\ Alkofer and C.\ Fischer.
CDR is grateful for support from the Special Research Centre for the
Subatomic Structure of Matter at the University of Adelaide, Australia, and
the hospitality of its staff during a visit in which some of this work was
completed.
This work was supported by: the Department of Energy, Nuclear Physics
Division, under contract no.\ \mbox{W-31-109-ENG-38}; the National Science
Foundation under contract nos.\ PHY-0071361 and INT-0129236; and benefited
from the resources of the National Energy Research Scientific Computing
Center.
\end{acknowledgments} 
%------------------------------------------------------------------------ 

\bibliography{aqlp}

\end{document}